\documentclass[a4paper,american]{article}
\usepackage[]{fontenc}
\usepackage[latin1]{inputenc}
\usepackage{amsmath}
\usepackage{graphicx}
\usepackage{amssymb}

\makeatletter
\usepackage{amsmath}
\usepackage{graphicx}
\usepackage{amssymb}
\usepackage{amsfonts}
\usepackage{amsmath}
\usepackage{amssymb}
\usepackage{graphicx}
\usepackage{babel}
\usepackage{babel}

\usepackage{babel}
\makeatother
\begin{document}

\title{Quantum Walk with a time-dependent coin}

\author{M.C. Bañuls$^{1}$, C. Navarrete$^{1}$, A. Pérez$^{1}$, Eugenio
Roldán$^{2}$, \and and J.C. Soriano$^{2}$\\
$^{1}$Departament de Física Teòrica and IFIC, \\
 Universitat de València-CSIC, \\
 Dr. Moliner 50, 46100--Burjassot, Spain\\
 $^{2}$Departament d'Òptica, Universitat de València, \\
 Dr. Moliner 50, 46100--Burjassot, Spain}

\maketitle
\begin{abstract}
We introduce quantum walks with a time-dependent coin, and show how
they include, as a particular case, the generalized quantum walk recently
studied by Wojcik et al. {[}Phys. Rev. Lett. \textbf{93}, 180601(2004){]}
which exhibits interesting dynamical localization and quasiperiodic
dynamics. Our proposal allows for a much easier implementation of
this particular rich dynamics than the original one. Moreover, it
allows for an additional control on the walk, which can be used to
compensate for phases appearing due to external interactions. To illustrate
its feasibility, we discuss an example using an optical cavity. We
also derive an approximated solution in the continuous limit (long--wavelength
approximation) which provides physical insight about the process. 
\end{abstract}

\section{Introduction}

Quantum walks (QWs) \cite{Aharonov93,Meyer96} constitute a promising
ingredient in the research of quantum algorithms \cite{Kempe03} but
have also an intrinsic interest, reinforced through their connection
with quantum cellular automata \cite{Meyer96} and with phenomena
such as Anderson localization or quantum chaos \cite{Buerschaper,Wojcik,Romanelli}.

Both in the discrete or continuous version, QWs provide a mean to
explore all possible paths on a lattice in a parallel way, which is
natural for quantum evolution, together with constructive quantum
interference along the paths. Thus they can allow the development
of probabilistic algorithms in a more efficient way than their classical
counterparts \cite{Farhi98}. It is therefore crucial to fully explore
the possibilities offered by QWs, especially in connection with their
physical implementation.

Modified QWs can give rise to new physical phenomena, along with more
efficient algorithmic applications. Different variations of the standard
discrete time QW have been proposed, including QWs with two entangled
particles \cite{Omar04} or entangled coins \cite{Venegas05}, multi--states
QWs \cite{Inui05a,Inui05b}, as well as QWs with alternation of different
quantum coins in a certain sequence~\cite{Ribeiro04}. More interesting
to us are the generalized QWs that modify the time evolution by the
acquisition of position--dependent phases by the walker at every step~\cite{Buerschaper,Wojcik,Romanelli}.
Those generalizations show phenomena that differ from the typical
linear spreading of the wave function in the standard QW, such as
quantum resonances and dynamic localization.

Within this spirit, we explore a modification on the standard coined
QW, which consists on the introduction of a time-dependent coin. As
we discuss, this modification introduces new possibilities on the
walk which are worth investigating. Here we concentrate on a particular
time dependent coin that leads to QW equations nearly identical to
those corresponding to the generalized QW introduced by Wojcik et
al. \cite{Wojcik} (see also \cite{Buerschaper,Romanelli}). Our approach
presents the advantage that the corresponding modifications are made
on the coin alone, which is a simple one-qubit system, in contrast
to the original proposal, which requires operations to be performed
on a large system (the Hilbert space of the walking particle). Moreover,
using a time-dependent coin can be used as a control mechanism to
compensate for a phase arising from some external influence. We illustrate
this idea with an example.

This paper is organized as follows. In Section 2 we introduce the
basic idea of a time-dependent coined QW and relate it to previous
works. In Section 3, we first review the main aspects of the generalized
QW introduced by Wojcik et al. \cite{Wojcik}, and then show how an
extra transformation (with respect to the standard coined QW) on the
walking particle, can be encoded into a time-dependent coin, and show
the equivalence between the obtained generalized QW and that of \cite{Wojcik}.
We also discuss the utility of a time-dependent coin as a control
mechanism. In Section 4, we show how this generalized QW could be
implemented in an optical cavity. Then, in Section 5, we derive an
approximated continuous limit, a long wave--length approximation to
this time-dependent QW, which is appropriate for describing dynamic
localization. Finally, in Section 6 we summarize our main results.

\section{Time-dependent coined walks}

The standard QW corresponds to the evolution on a one-dimensional
lattice of a quantum system (the walker) coupled to a bidimensional
system (the coin), under repeated application of a pair of discrete
operators. Let $\mathcal{H}_{P}$ be the Hilbert space of the walker,
with $\left\{ \left|n\right\rangle ,n\in\mathbb{Z}\right\} $ a basis
of $\mathcal{H}_{P}$; and let $\mathcal{H}_{C}$ be the Hilbert space
of the coin, with basis $\left\{ \left|u\right\rangle ,\left|d\right\rangle \right\} $.
The state of the total system belongs to the space $\mathcal{H}=\mathcal{H}_{C}\otimes\mathcal{H}_{P}$
and, at a given time, can be expressed as \begin{equation}
\left|\psi\left(t\right)\right\rangle =\sum_{n}\left[u_{n}\left(t\right)\left|n,u\right\rangle +d_{n}\left(t\right)\left|n,d\right\rangle \right].\end{equation}
 The evolution of the system is governed by two operators: (i) an
arbitrary unitary transformation $\hat{C}$ acting on $\mathcal{H}_{C}$,
which can be any unitary 2x2 matrix and is usually chosen as\begin{equation}
\hat{C}=\left(\begin{array}{cc}
\sqrt{\rho} & \sqrt{1-\rho}\\
\sqrt{1-\rho} & -\sqrt{\rho}\end{array}\right)\label{Cstandard}\end{equation}
 (with $\rho=1/2$ the balanced Hadamard coin $H=\left(\sigma_{x}+\sigma_{z}\right)/\sqrt{2}$
is recovered); and (ii) the conditional displacement operator $\hat{S}$
acting on $\mathcal{H}_{P}$\begin{eqnarray}
\hat{S}\left|n,u\right\rangle  & = & \left|n+1,u\right\rangle ,\label{S1}\\
\hat{S}\left|n,d\right\rangle  & = & \left|n-1,d\right\rangle .\label{S2}\end{eqnarray}
 Altogether, they produce the evolution from instant $t-1$ to $t$
given by \begin{equation}
\left|\psi\left(t\right)\right\rangle =\hat{S}\hat{C}\left|\psi\left(t-1\right)\right\rangle .\label{defcoint0}\end{equation}

In this paper, we introduce the idea of a modified QW, where the coin
changes during the evolution, i.e. $\hat{C}(t)$. In this case, the
evolution from instant $t-1$ to $t$ is defined by\begin{equation}
\left|\psi\left(t\right)\right\rangle =\hat{S}\hat{C}(t)\left|\psi\left(t-1\right)\right\rangle \label{defcoint}\end{equation}

A particular case of this would be the proposal in~\cite{Ribeiro04},
in which two fixed standard coins were alternated in a given sequence,
leading to a sub-ballistic wave-function spreading for some particular
choices of the coin series. In order to be more specific, we study
the effect of a time-dependent coin of the special form \begin{equation}
\hat{C}(t)=\left(\begin{array}{cc}
\sqrt{\rho}e^{-i\Phi(t)} & \sqrt{1-\rho}e^{-i\Phi(t)}\\
\sqrt{1-\rho}e^{i\Phi(t)} & -\sqrt{\rho}e^{i\Phi(t)}\end{array}\right).\label{Cgral}\end{equation}
 Notice that (\ref{Cgral}) can be obtained as the sequence of two
operations, i.e., \begin{equation}
\hat{C}(t)=\hat{C}_{0}(t)\hat{C},\end{equation}
 with\begin{equation}
\hat{C}_{0}(t)=\left(\begin{array}{cc}
e^{-i\Phi(t)} & 0\\
0 & e^{i\Phi(t)}\end{array}\right),\end{equation}
 and $\hat{C}$ given by (\ref{Cstandard}). 

Again, $\Phi(t)$ is quite a general function. In this article we
shall restrict ourselves to a particular case that, as commented above,
leads to a generalized QW which is nearly identical to that analyzed
in Ref. \cite{Wojcik}. Other possibilities will be considered in
a future work.

\section{Using time-dependent coins to implement dynamic localization and
quasiperiodic dynamics}

Recently, Wojcik et al.~\cite{Wojcik} (see also \cite{Buerschaper,Romanelli})
showed that a generalization of the QW (GQW in the following) in which
a position-dependent phase $\Phi\left(n\right)\propto n$ was acquired
by the walker with each evolution step, produces quasiperiodic dynamics
and localization effects. There is a physical reason for introducing
$\Phi\left(n\right)$: the walker is a physical system that evolves
in time, and this evolution can introduce such phases, via e.g. external
interactions.

Here we show that such generalization can in fact be recast as a QW
with a time-dependent coin. We concentrate here on the GQW of \cite{Wojcik},
which is equivalent to that of \cite{Buerschaper}, but our approach
can easily be shown to cover also Romanelli's et al. proposal \cite{Romanelli}.
In fact, the only difference in the dynamical equations, with respect
to \cite{Buerschaper,Wojcik}, is that the position dependent phase
in \cite{Romanelli} goes like $\Phi\left(n\right)\propto n^{2}$.

\subsection{GQW}

Let us briefly present the GQW introduced in \cite{Wojcik}. We first
define the discrete position operator $\hat{n}$ \ such that \begin{equation}
\hat{n}\left|n\right\rangle =n\left|n\right\rangle ,\end{equation}
and, related to this one, the phase operator \begin{equation}
\hat{E}_{0}\equiv e^{i\phi_{0}\hat{n}}\label{E0}\end{equation}
where $\phi_{0}$ is a constant. Following \cite{Wojcik}, the evolution
of the system is governed by \begin{equation}
\left|\bar{\psi}\left(t\right)\right\rangle =\hat{S}\hat{C}\hat{E}_{0}\left|\bar{\psi}\left(t-1\right)\right\rangle .\label{evolW}\end{equation}
The state of the system at a given time can be expressed as \begin{equation}
\left|\bar{\psi}\left(t\right)\right\rangle =\sum_{n}\left[\bar{u}_{n}\left(t\right)\left|n,u\right\rangle +\bar{d}_{n}\left(t\right)\left|n,d\right\rangle \right],\label{expansion}\end{equation}
from which it is easy to obtain \begin{eqnarray}
\bar{u}_{n}\left(t\right) & = & e^{i\left(n-1\right)\phi_{0}}\left[\sqrt{\rho}\,\,\bar{u}_{n-1}\left(t-1\right)+\sqrt{1-\rho}\,\,\bar{d}_{n-1}\left(t-1\right)\right],\label{W1}\\
\bar{d}_{n}\left(t\right) & = & e^{i\left(n+1\right)\phi_{0}}\left[\sqrt{1-\rho}\,\,\bar{u}_{n+1}\left(t-1\right)-\sqrt{\rho}\,\,\bar{d}_{n+1}\left(t-1\right)\right].\label{W2}\end{eqnarray}
From the solution of these equations one can evaluate the probability
of finding the walker at the lattice point $n$\ at iteration $t$\ by
using\begin{equation}
\mathcal{P}_{n}(t)=\left|\bar{u}_{n}\left(t\right)\right|^{2}+\left|\bar{d}_{n}\left(t\right)\right|^{2}.\end{equation}

We now briefly summarize the main features of the solutions of Eqs.
(\ref{W1},\ref{W2}). Wojcik et al. \cite{Wojcik} found that for
rational values of $\phi_{0}/2\pi$ dynamical localization, shown
by a 'quasiperiodic' behavior of the standard deviation $\sigma$
of the probability distribution, is observed during a transient regime,
but for long enough times a ballistic diffusion occurs. For irrational
values of $\phi_{0}/2\pi$, on the contrary, the diffusion becomes
suppressed, and the walk shows dynamic localization around the starting
point for arbitrarily long $t$. Let us consider the case of a rational
$\phi_{0}/2\pi$ in more detail.

First we notice that the probability distribution $\mathcal{P}_{n}(t)$
is invariant under the change\begin{equation}
\phi_{o}\rightarrow\phi_{o}+\pi k\text{, with }k\in\mathbb{Z}.\label{Symmetry}\end{equation}
Then, if we focus on rational values of $\phi_{0}/2\pi$, we can restrict
the study to\begin{equation}
\phi_{o}=2\pi\frac{q}{p},\,\,\,\, and\,\,\,\,\,\,0\leq\frac{q}{p}<\frac{1}{2},\end{equation}
 where $q/p$ is an irreducible fraction. Moreover, the study can
be limited to even values of $p$, since given a case $q/p$ with
odd $p$, there is a value $\left(2q-p\right)/\left(2p\right)$ with
even denominator leading to the same probability distribution, as
a consequence of symmetry (\ref{Symmetry}).

\begin{figure}
\includegraphics[%
  scale=0.5]{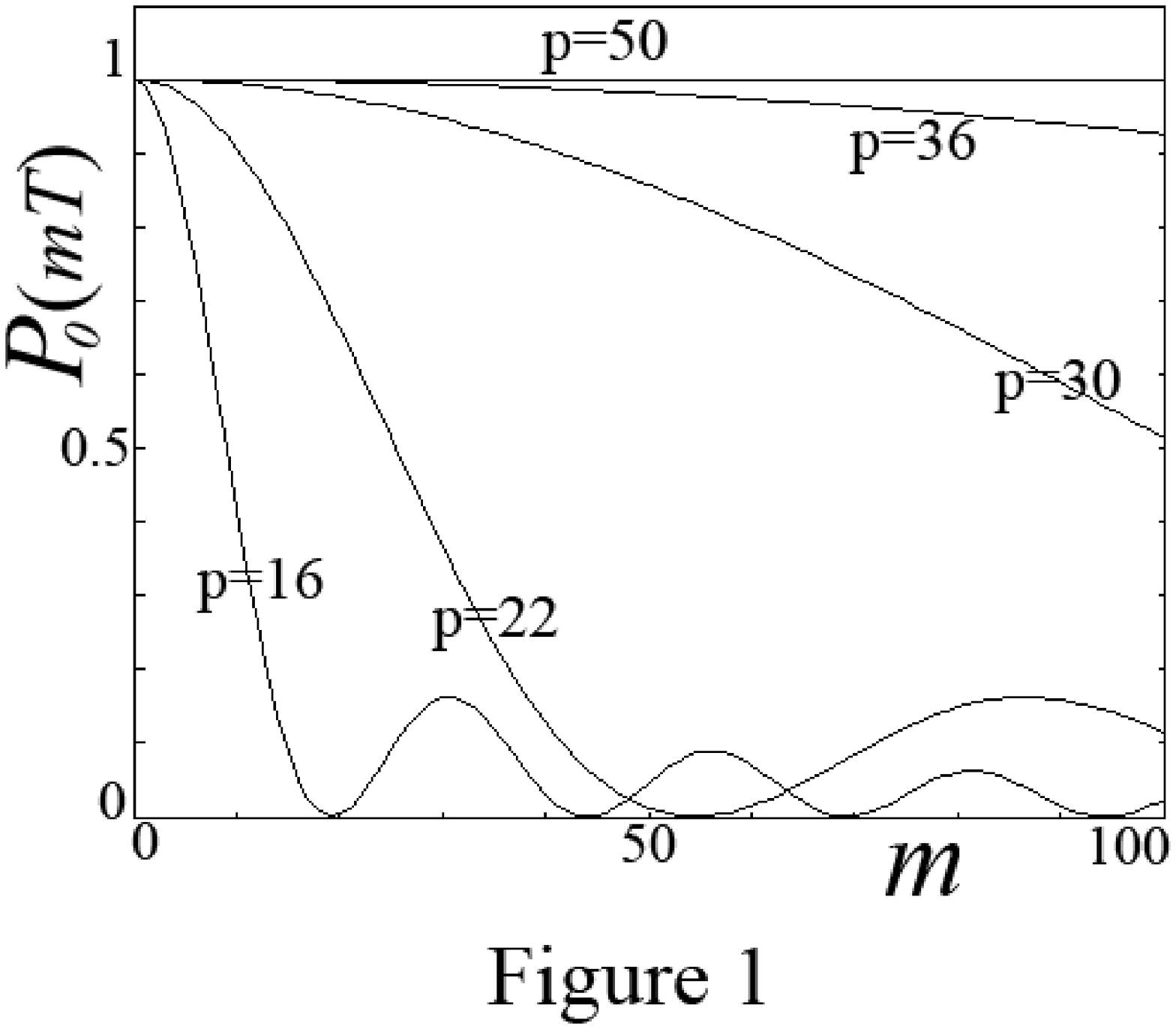}

\caption{This figure illustrates the localization features of the GQW during
the transient regime, by showing the probability that the walker returns
close to its initial position after $m$ quasiperiods. As can be readily
observed, the number of these quasiperiods before the transient ends
increases with $p$.}
\end{figure}

Keeping this in mind, a numerical analysis of Eqs. (\ref{W1},\ref{W2})
shows that, given an even $p$, the solution of the GQW shows a quasiperiod
$T=p$ during the above-mentioned transient regime. The duration of
this transient, i.e., the number of quasiperiods that it exhibits,
turns out to be larger the larger is $p$. In other words, the probability
that the walker returns to the initial position after one quasiperiod,
($P_{0}\left(mT\right)\simeq1$) with $m\in\mathbb{N}$, increases
with $p$, as we show in Fig. 1. This figure clearly shows that the
loss of localization takes place more slowly for larger values of
$p$. %
\begin{figure}
\includegraphics[%
  scale=0.3]{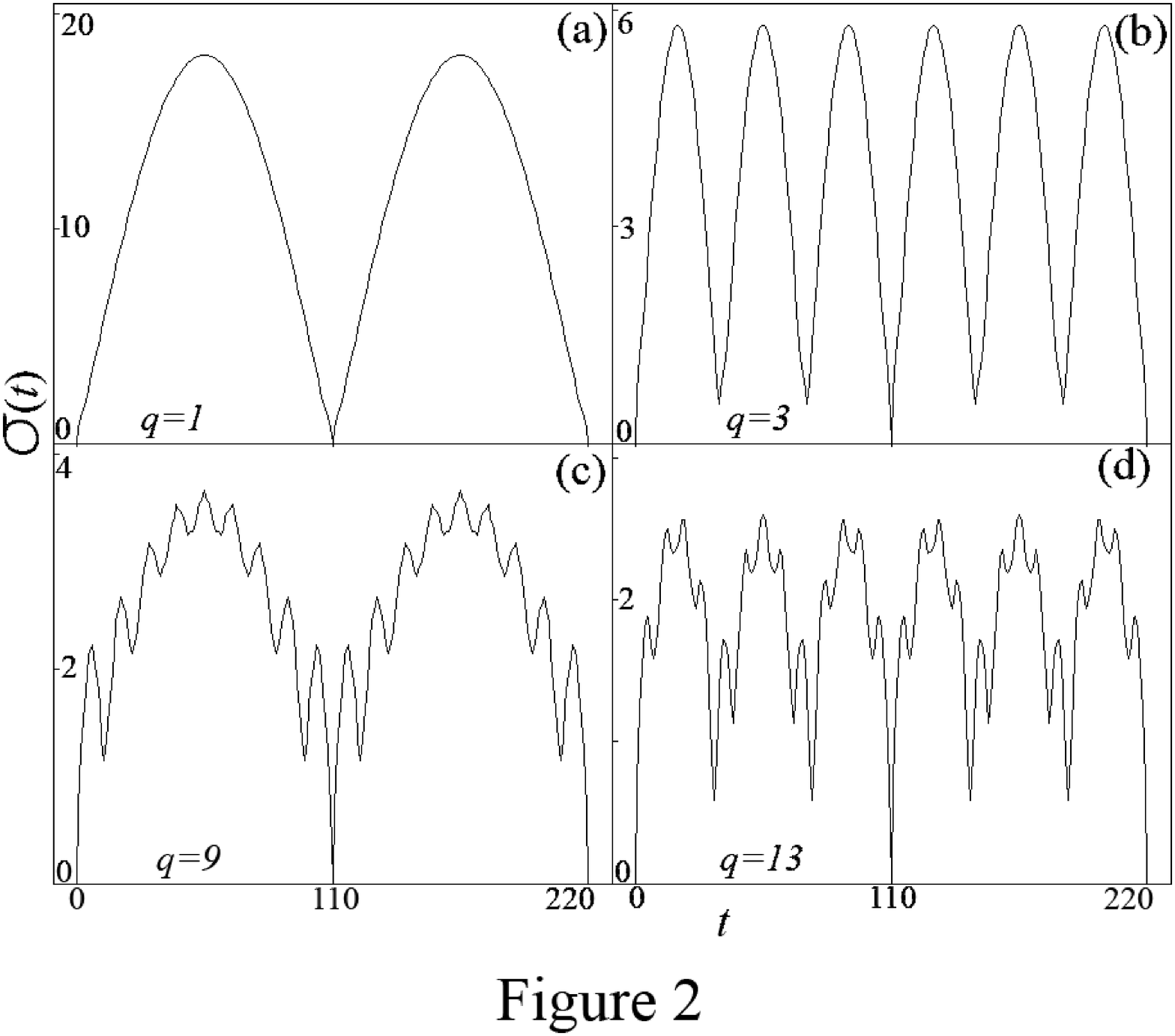}

\caption{Here we show how changing the value of $q$ influences the behavior
of the walker. We have chosen symmetric initial conditions $u_{n}(0)=\delta_{0,n}/\sqrt{2}$
and $d_{n}(0)=i\delta_{0,n}/\sqrt{2}$, the Hadamard coin ($\rho=1/2$),
$p=110$, and four different values of $q$. The standard deviation
$\sigma(t)$ against time has exactly $q$ peaks within one quasiperiod.}
\end{figure}

Apart from this oscillation of quasiperiod $p$, the standard deviation
$\sigma$ of the probability distribution shows a faster secondary
oscillation that depends on $q$. In fact, one finds $q$ secondary
oscillations within each period of the main oscillation. These secondary
oscillations are more pronounced the smaller is $q$ and the larger
is $T$. This is clearly appreciated in Fig. 2, where we show the
evolution of $\sigma$ for $T=p=110$ and four different values of
$q$. Notice how the GQW returns (only approximately, remind this
is a transient behavior) to the initial condition, $\sigma\left(t=0\right)=0$,
when $t=mT$, oscillating $q$ times between $t=mT$ and $t=\left(m+1\right)T$.

\subsection{An alternative approach}

Here we present our alternative approach. Let us define \begin{equation}
\hat{C}_{0}=\left(\begin{array}{cc}
e^{-i\phi_{0}} & 0\\
0 & e^{i\phi_{0}}\end{array}\right).\label{C0}\end{equation}
It is straightforward to show that the following relationship holds
\begin{equation}
\hat{S}\hat{E}_{0}=\hat{C}_{0}\hat{E}_{0}\hat{S},\end{equation}
with $\hat{S}$ given by Eqs.(\ref{S1},\ref{S2}) and $\hat{E}_{0}$
given by Eq. (\ref{E0}).

Repeated use of the above expression, together with the evolution
Eq. (\ref{evolW}), leads to a modified form of the evolution equation,
which can be expressed as \begin{equation}
\left|\bar{\psi}\left(t\right)\right\rangle =\left(\hat{E}_{0}\right)^{t}\left|\psi\left(t\right)\right\rangle ,\label{connection}\end{equation}
where $\left|\psi\left(t\right)\right\rangle $ verifies \begin{eqnarray}
\left|\psi\left(t\right)\right\rangle  & = & \hat{S}\hat{C}(t)\left|\psi\left(t-1\right)\right\rangle ,\\
\left|\psi\left(0\right)\right\rangle  & = & \left|\bar{\psi}\left(0\right)\right\rangle ,\end{eqnarray}
and $\hat{C}(t)$ is a time-dependent coin operator, defined as \begin{equation}
\hat{C}(t)\equiv\left(\hat{C}_{0}\right)^{t}\hat{C}=\left(\begin{array}{cc}
\sqrt{\rho}e^{-i\phi_{0}t} & \sqrt{1-\rho}e^{-i\phi_{0}t}\\
\sqrt{1-\rho}e^{i\phi_{0}t} & -\sqrt{\rho}e^{i\phi_{0}t}\end{array}\right),\label{Cnostra}\end{equation}
i.e., $\left(\hat{C}_{0}\right)^{t}=\hat{C}_{0}(t)$, c.f. Eq. (\ref{Cgral}),
with $\Phi(t)=\phi_{0}t$. By this simple procedure we have demonstrated
that the generalization of the QW introduced in Ref. \cite{Wojcik}
can be obtained by introducing a suitable time dependent coin. Although
the probability amplitudes are not identical to those of the GQW due
to the phase factors appearing in Eq. (\ref{connection}), the probability
distributions are the same obtained either with $\left|\bar{\psi}\left(t\right)\right\rangle $
or with $\left|\psi\left(t\right)\right\rangle $, and both descriptions
are thus equivalent from this point of view.

The equivalence nevertheless breaks down for the QW on the circle,
as the phase added in \cite{Wojcik} depends on the position, so that
a difference may arise in the circle when passing from position $-L$
to position $+L$.

Now we write down explicitly the equations of evolution for our alternative
approach. By performing a decomposition analogous to Eq. (\ref{expansion}),
the equations of evolution become \begin{eqnarray}
u_{n}\left(t\right) & = & e^{-it\phi_{0}}\left[\sqrt{\rho}u_{n-1}\left(t-1\right)+\sqrt{1-\rho}d_{n-1}\left(t-1\right)\right],\label{V1}\\
d_{n}\left(t\right) & = & e^{it\phi_{0}}\left[\sqrt{1-\rho}u_{n+1}\left(t-1\right)-\sqrt{\rho}d_{n+1}\left(t-1\right)\right].\label{V2}\end{eqnarray}
 which are equivalent to Eqs.(\ref{W1},\ref{W2}), as Eq. (\ref{connection})
provides the connection between both descriptions, which reads\begin{equation}
\bar{u}_{n}\left(t\right)=u_{n}\left(t\right)e^{int\phi_{0}},\,\,\,\,\,\,\,\,\bar{d}_{n}\left(t\right)=d_{n}\left(t\right)e^{int\phi_{0}}.\label{reluds}\end{equation}

We now transform the coupled equations (\ref{V1}) and (\ref{V2})
into space-time recursive equations for $u_{n}$ and $d_{n}$, where
both components are decoupled. We start from \begin{eqnarray}
\left|\psi\left(t+1\right)\right\rangle  & = & \hat{S}\hat{C}(t+1)\left|\psi\left(t\right)\right\rangle ,\\
\left|\psi\left(t-1\right)\right\rangle  & = & \hat{C}^{\dagger}(t)\hat{S}^{\dagger}\left|\psi\left(t\right)\right\rangle ,\end{eqnarray}
 and making use of \begin{equation}
\hat{C}(t+1)=C_{0}\hat{C}(t),\end{equation}
 we obtain, after some algebra \begin{equation}
C_{0}^{\dagger}\left|\psi\left(t+1\right)\right\rangle -\left|\psi\left(t-1\right)\right\rangle =\sqrt{\rho}\sum_{a=u,d}\sum_{n}\left[a_{n-1}\left(t\right)e^{-it\phi_{0}}-a_{n+1}\left(t\right)e^{it\phi_{0}}\right]\left|n,a\right\rangle \end{equation}
 or, equivalently, \begin{eqnarray}
u_{n}\left(t+1\right)e^{i\phi_{0}}-u_{n}\left(t-1\right) & = & \sqrt{\rho}\left[u_{n-1}\left(t\right)e^{-it\phi_{0}}-u_{n+1}\left(t\right)e^{it\phi_{0}}\right],\label{V3}\\
d_{n}\left(t+1\right)e^{-i\phi_{0}}-d_{n}\left(t-1\right) & = & \sqrt{\rho}\left[d_{n-1}\left(t\right)e^{-it\phi_{0}}-d_{n+1}\left(t\right)e^{it\phi_{0}}\right],\label{V4}\end{eqnarray}
 Finally, the probability of finding the walker at the lattice point
$n$ at iteration $t$ is given by\begin{equation}
\mathcal{P}_{n}(t)=\left|u_{n}\left(t\right)\right|^{2}+\left|d_{n}\left(t\right)\right|^{2}\equiv\mathcal{P}_{n}^{u}(t)+\mathcal{P}_{n}^{d}(t).\label{exprob}\end{equation}
Since $\left|u_{n}\left(t\right)\right|=\left|\bar{u}_{n}\left(t\right)\right|$
and $\left|d_{n}\left(t\right)\right|=\left|\bar{d}_{n}\left(t\right)\right|$,
there is no difference between the probability distribution for the
QW on a line calculated with Eqs. (\ref{V1}, \ref{V2}) or with Eqs.
(\ref{W1}, \ref{W2}), as already commented.

\subsection{The time-dependent coin as a control mechanism}

In this subsection we discuss how a time-dependent coin can be used
to gain control over a possible phase arising during the walk, as
a consequence of additional interactions \cite{Buerschaper}. We illustrate
this idea with an example which shows that, at least in some cases,
the position-depending phase acquired between two steps in the walk
could be canceled by an appropriate action on the coin.

Let us assume that the walker is subjected to the effect of the GQW
defined by Eq. (\ref{evolW}). We have shown in the previous subsection
that this kind of QW is equivalent (modulo a final phase) to one with
a time-dependent coin. Intuitively, if one wants to compensate for
the phases acquired during the GQW, one should replace the coin operator
$\hat{C}$ by a time-dependent operator $\hat{C}(t)$ defined by $\hat{C}(t)=\left(\hat{C}_{0}^{\dagger}\right)^{t}\hat{C}$.
In this way, the evolution is governed by\[
\left|\psi\left(t\right)\right\rangle =\hat{S}\hat{C}(t)E_{0}\left|\psi\left(t-1\right)\right\rangle .\]
 Using the properties given in section 3.2, one obtains:

\begin{equation}
\left|\psi\left(t\right)\right\rangle =\left(\hat{E}_{0}\right)^{t}\left(\hat{S}\hat{C}\right)^{t}\left|\psi\left(0\right)\right\rangle ,\label{control}\end{equation}
 Showing that the combined action of the phase operator $E_{0}$ and
the time-dependent coin defined above, is equivalent (up to a phase
give by the action of $\left(\hat{E}_{0}\right)^{t}$) to the standard
quantum walk introduced in section 2. In other words, Eq. (\ref{control})
can be written, when decomposed in the $|n,u>,|n,d>$ basis, as \[
a_{n}(t)=e^{int\phi_{0}}a_{n}^{s}(t)\]
where $a=u,d$. The coefficients $a_{n}(t)$ then correspond to Eq.
(\ref{control}), whereas $a_{n}^{s}(t)$ stand for the standard QW.
In this way, the complex dynamics arising from the GQW translates
into a trivial phase.

\section{Implementing the generalized quantum walk}

Along recent years there have been many proposals for the experimental
implementation of QWs. These cover both systems whose dynamics can
be described only within the framework of quantum mechanics \cite{Travaglione,Dur,Sanders,Zhao,Di04,Eckert05,Agarwal}
as well as setups whose description does not require quantum mechanics
\cite{Hillery,Knight03,Knight03(b),Jeong,Roldan05}. In fact, the
QW on the line was nearly implemented in an optical cavity \cite{Bouwmeester},
as it was highlighted in \cite{Knight03} and fully discussed in \cite{Knight03(b)}.
Although an experimental realization of the QW using only classical
means has been communicated recently \cite{Do05}, it is a fact that
there has been little experimental research about this process.

Here we comment on how the GQW we are studying could be implemented
in an optical cavity. We follow our approach to the GQW as it is more
easily implementable than the original proposal by Wojcik et al. \cite{Wojcik}.
This is due to the fact that with our approach it is only needed to
modify the unitary transformation acting on the qubit, which is a
2--dimensional system, while the original proposal \cite{Wojcik}
implies acting on all the points of the lattice.

In \cite{Knight03,Knight03(b)}, it was shown that the QW on the line
can be implemented by the frequency of a quasi--monochromatic field,
e.g. an optical pulse of appropriate duration, inside an optical cavity.
As stated, in this classical implementation the role of the walker
is played by the field frequency, and the role of the coin can be
played, e.g., by the field polarization. The simplest scheme is that
represented in Fig. 3 \cite{Knight03}, without EOMbis (see below
for the role of this element): The electrooptic modulator (EOM) implements
the displacement operator, Eqs. (\ref{S1}, \ref{S2}), by increasing
(decreasing) the frequency of the horizontal (vertical) polarization
component of the field. As for the unitary transformation, $\hat{C}$,
it is performed by a half--wave plate (HWP) with suitably oriented
fast axis \cite{Spreeuw01}. Thus in a cavity round-trip a step of
the QW is performed. The optical cavity allows the repetition of the
process through feedback. The number of steps of the QW that can be
implemented depends on factors such as the technical limitations of
the EOM and on the losses of the cavity (this last factor could be
compensated by introducing gain in the cavity, as in \cite{Bouwmeester}).
We address the interested reader to \cite{Knight03(b)} for more details.
Let us remark that this simple scheme is very close to what was actually
performed in the experiment of Bouwmeester et al. \cite{Bouwmeester}
(see \cite{Knight03(b)} for a full discussion). %
\begin{figure}
\includegraphics[%
  scale=0.25]{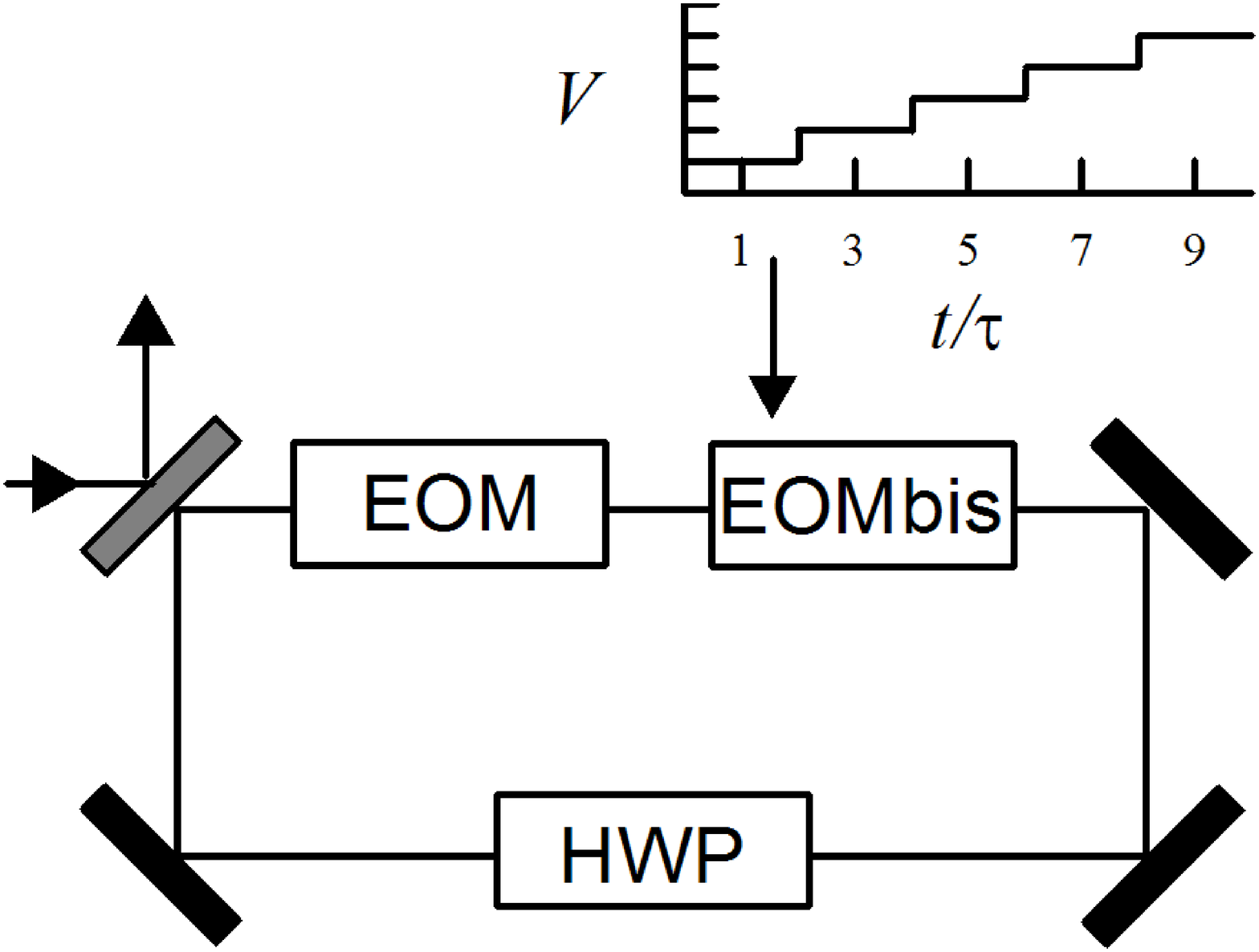}

\caption{The proposed experimental setup for implementing the GQW. See the
text for the description of the different components. The upper part
represents the staircase voltage that has to be applied to the EOMbis
to implement the time-dependent coin, and $\tau$ is the cavity roundtrip
time. }
\end{figure}

We can take this scheme as a basis for the implementation of the generalized
QW. In order to perform the GQW, one needs to implement the time dependent
unitary transformation $\hat{C}\left(t\right)$, Eq. (\ref{Cnostra}).
This can be done by adding one optical element between EOM and HWP
to implement $\left(\hat{C}_{0}\right)^{t}$, Eq. (\ref{C0}). This
is the role played by EOMbis in Fig. 3. Consider first a single step
of the GQW, i.e., that corresponding to iteration $t$. For this $t$
one must implement $\left(\hat{C}_{0}\right)^{t}$, which can be done
in a straightforward way: The implementation simply consists in the
addition (subtraction) of $\phi_{0}t$ to the phase of the horizontal
(vertical) polarization component of the field. This can be easily
carried out, e.g., by introducing a second EOM, EOMbis in Fig. 3,
to which a suitable (constant) voltage is applied. Now, in order to
implement $\hat{C}\left(t\right)$, this added (subtracted) phase
must be increased at each cavity round-trip, what is done by applying
a staircase voltage to EOMbis (represented in Fig. 3): The voltage
must remain constant while the light pulse is traversing EOMbis, in
order to modify the phase and not the field frequency, and then be
rapidly increased for the phase increment takes the value $\phi_{0}\left(t+1\right)$
in the subsequent round-trip. We think that this simple scheme, which
can be implemented with current technology (it consists in adding
a single element to the device already used in \cite{Bouwmeester}),
could allow the experimental investigation of the GQW.

\section{A long--wavelength approximation to the generalized quantum walk}

Up to now we have shown how the GQW can be alternatively produced
by means of a time dependent coin, how it could be experimentally
implemented, and also how the time dependent coin can be used in the
presence of phases in the walker displacement for controlling or tuning
the GQW. Now we will try to get some insight into the physics of the
GQW, by deriving a continuous version through a long--wavelength approximation.
In this way we derive a wave equation that constitutes a continuous
propagation analog of the GQW. The analogy helps to visualize the
kind of physical process that produces the GQW.

We introduce here a long wave approximation by following the same
lines as in \cite{Knight03,Knight04}. Our starting point is the recurrence
equation \begin{equation}
a_{n}\left(t+1\right)-a_{n}\left(t-1\right)=\sqrt{\rho}\left[a_{n-1}\left(t\right)e^{-i\phi_{0}t}-a_{n+1}\left(t\right)e^{i\phi_{0}t}\right],\label{rec}\end{equation}
where $a_{m}\left(t\right)$ stands for both $u_{m}\left(t\right)$
and $d_{m}\left(t\right)$. In this way, Eq. (\ref{rec}) corresponds
to Eqs. (\ref{V3}) and (\ref{V4}) after the factors $e^{\pm i\phi_{0}}$
on the left hand side have been neglected %
\footnote{This approximation is perfectly justified. Perhaps it is more clearly
seen if instead of the unitary transformation (\ref{Cnostra}), one
uses \[
\hat{C}(t)=\left(\begin{array}{cc}
\sqrt{\rho}e^{i\phi_{0}\left(t-\frac{1}{2}\right)} & \sqrt{1-\rho}e^{i\phi_{0}\left(t-\frac{1}{2}\right)}\\
\sqrt{1-\rho}e^{-i\phi_{0}\left(t+\frac{1}{2}\right)} & -\sqrt{\rho}e^{-i\phi_{0}\left(t+\frac{1}{2}\right)}\end{array}\right),\]
In this case, the exponential factors we are neglecting do not appear
on the left-hand side, as in Eqs. (\ref{V3}, \ref{V4}), but on the
right--hand side of these equations in the form $e^{\pm i\phi_{0}\left(t\mp\frac{1}{2}\right)}$
which can be approximated by $e^{\pm i\phi_{0}t}$ for large enough
$t$.%
}. In \cite{Knight03,Knight04} it was shown that it is necessary to
introduce two discrete fields $A_{n}^{\pm}\left(t\right)$ in order
to preserve the symmetry of the QW. Thus we define the new fields
$A_{n}^{\pm}\left(t\right)$ through \begin{equation}
a_{n}\left(t\right)=A_{n}^{+}\left(t\right)+\left(-1\right)^{t}A_{n}^{-}\left(t\right).\label{plusminus}\end{equation}
By inserting this definition into Eq. (\ref{rec}), one immediately
obtains\begin{equation}
A_{n}^{\pm}\left(t+1\right)-A_{n}^{\pm}\left(t-1\right)=\pm\sqrt{\rho}\left[A_{n-1}^{\pm}\left(t\right)e^{-i\phi_{0}t}-A_{n+1}^{\pm}\left(t\right)e^{i\phi_{0}t}\right],\end{equation}
which is convenient to rewrite in the form

\begin{eqnarray}
A_{n}^{\pm}\left(t+1\right)-A_{n}^{\pm}\left(t-1\right) & = & \pm\sqrt{\rho}\left[A_{n-1}^{\pm}\left(t\right)-A_{n+1}^{\pm}\left(t\right)\right]\cos\phi_{0}t\notag\\
 &  & \mp i\sqrt{\rho}\left[A_{n-1}^{\pm}\left(t\right)+A_{n+1}^{\pm}\left(t\right)\right]\sin\phi_{0}t.\label{discrete}\end{eqnarray}
Denoting by $\bar{x}$ and $\bar{t}$ the continuous space and time
variables, and by $\Delta\bar{x}$ and $\Delta\bar{t}$ the spacing
between lattice points and time between iterations, we can define
the adimensional continuous variables $\xi=\bar{x}/\Delta\bar{x}$
and $\tau=$ $\bar{t}/$ $\Delta\bar{t}$ and think of Eq. (\ref{discrete})
as the discretization of the following partial differential equation
\begin{eqnarray}
\sum_{k=0}^{\infty}\frac{1}{\left(2k+1\right)!}\frac{\partial^{2k+1}}{\partial\tau^{2k+1}}A^{\pm}\left(\xi,\tau\right) & = & \mp\sqrt{\rho}\cos\left(\phi_{0}\tau\right)\sum_{k=0}^{\infty}\frac{1}{\left(2k+1\right)!}\frac{\partial^{2k+1}}{\partial\xi^{2k+1}}A^{\pm}\left(\xi,\tau\right)\notag\\
 &  & \mp i\sqrt{\rho}\sin\left(\phi_{0}\tau\right)\sum_{k=0}^{\infty}\frac{1}{\left(2k\right)!}\frac{\partial^{2k}}{\partial\xi^{2k}}A^{\pm}\left(\xi,\tau\right),\label{todos}\end{eqnarray}
which constitutes a continuous limit of the GQW.

Taking into account Eq. (\ref{plusminus}), and the fact that the
discrete fields $a_{n}\left(t\right)$ describe both $u_{n}\left(t\right)$
and $d_{n}\left(t\right)$, the continuous versions of these fields,
which we denote by $u\left(\xi,\tau\right)$ and $d\left(\xi,\tau\right)$,
are calculated through\begin{eqnarray}
u\left(\xi,\tau\right) & = & U^{+}\left(\xi,\tau\right)+\left(-1\right)^{t}U^{-}\left(\xi,\tau\right),\\
d\left(\xi,\tau\right) & = & D^{+}\left(\xi,\tau\right)+\left(-1\right)^{t}D^{-}\left(\xi,\tau\right),\end{eqnarray}
with $U^{\pm}\left(\xi,\tau\right)$ and $D^{\pm}\left(\xi,\tau\right)$
the solutions of Eq. (\ref{todos}) for $A^{\pm}\left(\xi,\tau\right)=U^{\pm}\left(\xi,\tau\right)$
and $A^{\pm}\left(\xi,\tau\right)=D^{\pm}\left(\xi,\tau\right)$,
respectively (see the Appendix).

The long-wavelength approximation consists in retaining the lowest
order in Eq. (\ref{todos}). Importantly, we further neglect the third
temporal derivative. We address the reader to the Appendix for full
details. After all of this, we are left with\begin{equation}
\frac{\partial}{\partial\tau}B^{\pm}\left(\xi,\tau\right)=\mp\sqrt{\rho}\left[\cos\left(\phi_{0}\tau\right)\frac{\partial}{\partial\xi}+\frac{i}{2}\sin\left(\phi_{0}\tau\right)\frac{\partial^{2}}{\partial\xi^{2}}+\frac{1}{6}\cos\left(\phi_{0}\tau\right)\frac{\partial^{3}}{\partial\xi^{3}}\right]B^{\pm}\left(\xi,\tau\right),\label{difer1}\end{equation}
where the new fields $B^{\pm}\left(\xi,\tau\right)$ defined by

\begin{equation}
B^{\pm}\left(\xi,\tau\right)=A^{\pm}(\xi,\tau)\exp\left[\mp i\frac{\sqrt{\rho}}{\phi_{0}}\cos(\phi_{0}\tau)\right],\label{change}\end{equation}
have been introduced. This equation can be solved analytically, and
the explicit solution is derived in the Appendix.

We notice that Eq. (\ref{difer1}) has time--periodic coefficients
and, consequently, as we have retained only the first derivative with
respect to time, their solutions are time--periodic. We can then expect
that the solutions of Eq. (\ref{difer1}) describe approximately the
periodic solutions of the GQW, which appear when $\phi_{0}$ is an
irrational multiple of $2\pi$, but not the quasiperiodic solutions
($\phi_{0}$ a rational multiple of $2\pi$) except in the cases with
very long quasiperiod. Obviously, this partial description of the
solutions is the price to be paid after neglecting the third time
derivative in Eq. (\ref{todos}).

Before discussing the physical meaning of Eq. (\ref{difer1}), let
us first compare the exact solution of the time-dependent coined QW,
Eqs. (\ref{V3},\ref{V4}), with the approximated continuous solution
we have just derived. In order to do that, we have chosen a value
for the phase $\phi_{0}$ ($\phi_{0}=$ $2\pi/150$) for which the
quasiperiod $T$ is very large ($T=$ $150$ in this case). We have
taken symmetrical initial conditions too (i.e. $u_{0}(0)=1/\sqrt{2}$
and $d_{0}(0)=i/\sqrt{2}$). For the continuous version, we will take
$A^{\pm}(\xi,0)$ to be a superposition of gaussians with a width
$w$ (see the Appendix for details). %
\begin{figure}
\includegraphics[%
  scale=0.35]{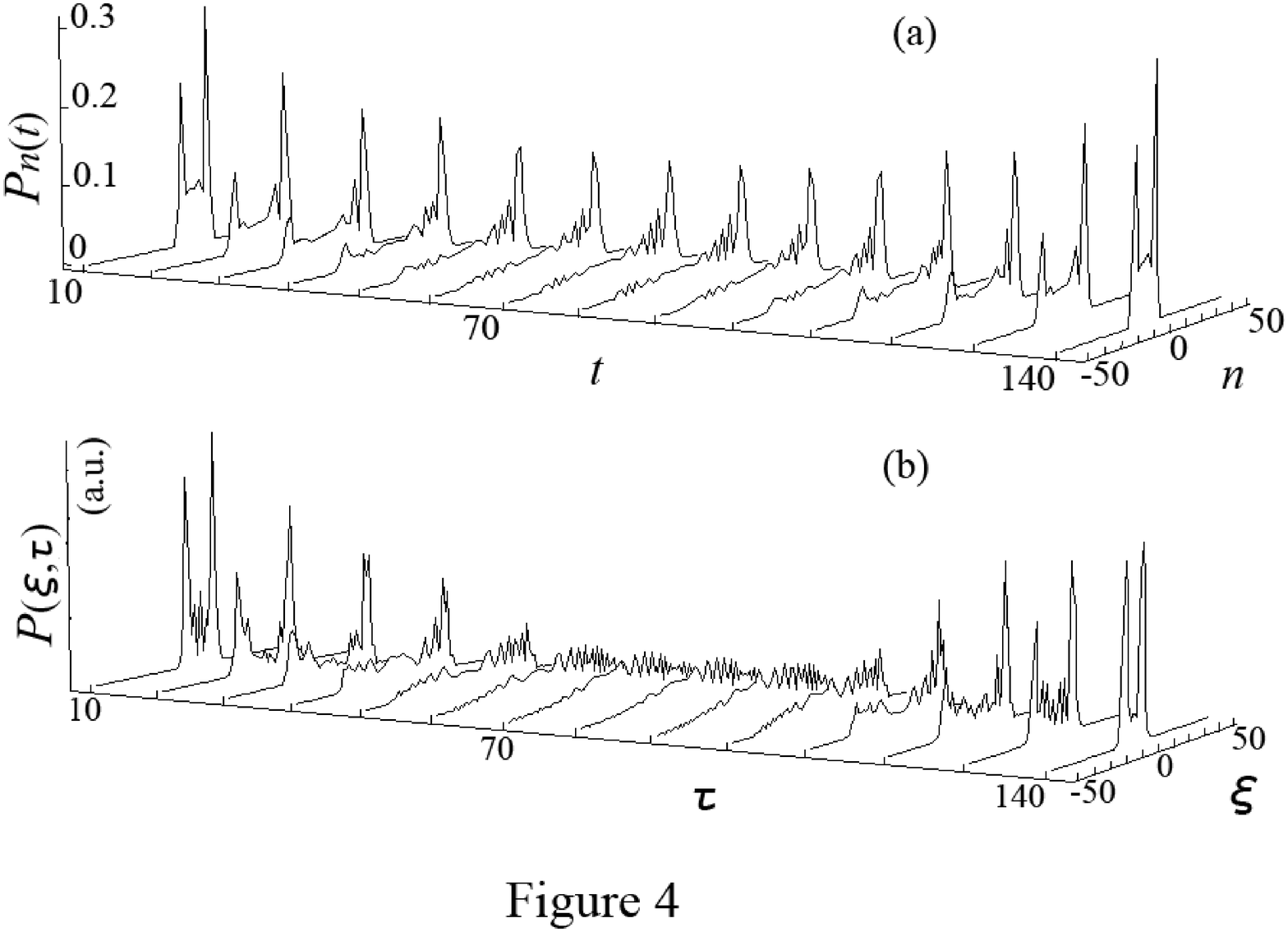}

\caption{A comparison of: (a) an exact numerical evaluation of Eqs. (\ref{V3},\ref{V4})
with (b), the long-wavelength approximation derived in this section.
Both calculations correspond to a value $\phi_{0}=$ $2\pi/150$.
Only even lattice points have been considered. }
\end{figure}

Fig. 4 shows both the exact probability distribution $\mathcal{P}_{n}(t)$,
Fig. 4(a), and the approximated continuous solution, Fig. 4(b), for
time running from $t=10$ to $t=140$, at intervals of $10$ time
units. For the exact probability distribution only even points of
the lattice, for which the probability is non zero, are shown and
joined for an easier visualization. We have chosen $w=0.65$ to evaluate
$\mathcal{P}(\xi,\tau)$. One sees how similar these distributions
are, except for $t$ close to $T/2$, where the continuous distribution
is wider. Then, the approximated continuous solution can be considered
as a good approximation for cases with periodic behavior or with quasiperiodic
behavior with very long transients.

Fig. 4 is complemented with Fig. 5, where we show: the exact $(n,\mathcal{P}_{n}(t))$
(with even points joined again) on the top row; the approximated $(\xi,\mathcal{P}(\xi,\tau))$
on the bottom row; and finally, in the middle row, the same as in
the bottom row (i.e., the long--wavelength approximation) but evaluated
only at discrete position values for a better comparison of the previous
two results. We do this for three different time values ($t=20$,
$t=70$, $t=110$). Again, as in Fig. 4, one sees how $\mathcal{P}_{n}(t)$
and $\mathcal{P}(\xi,\tau)$ are very similar, except near the semiperiod.
\begin{figure}
\includegraphics[%
  scale=0.3]{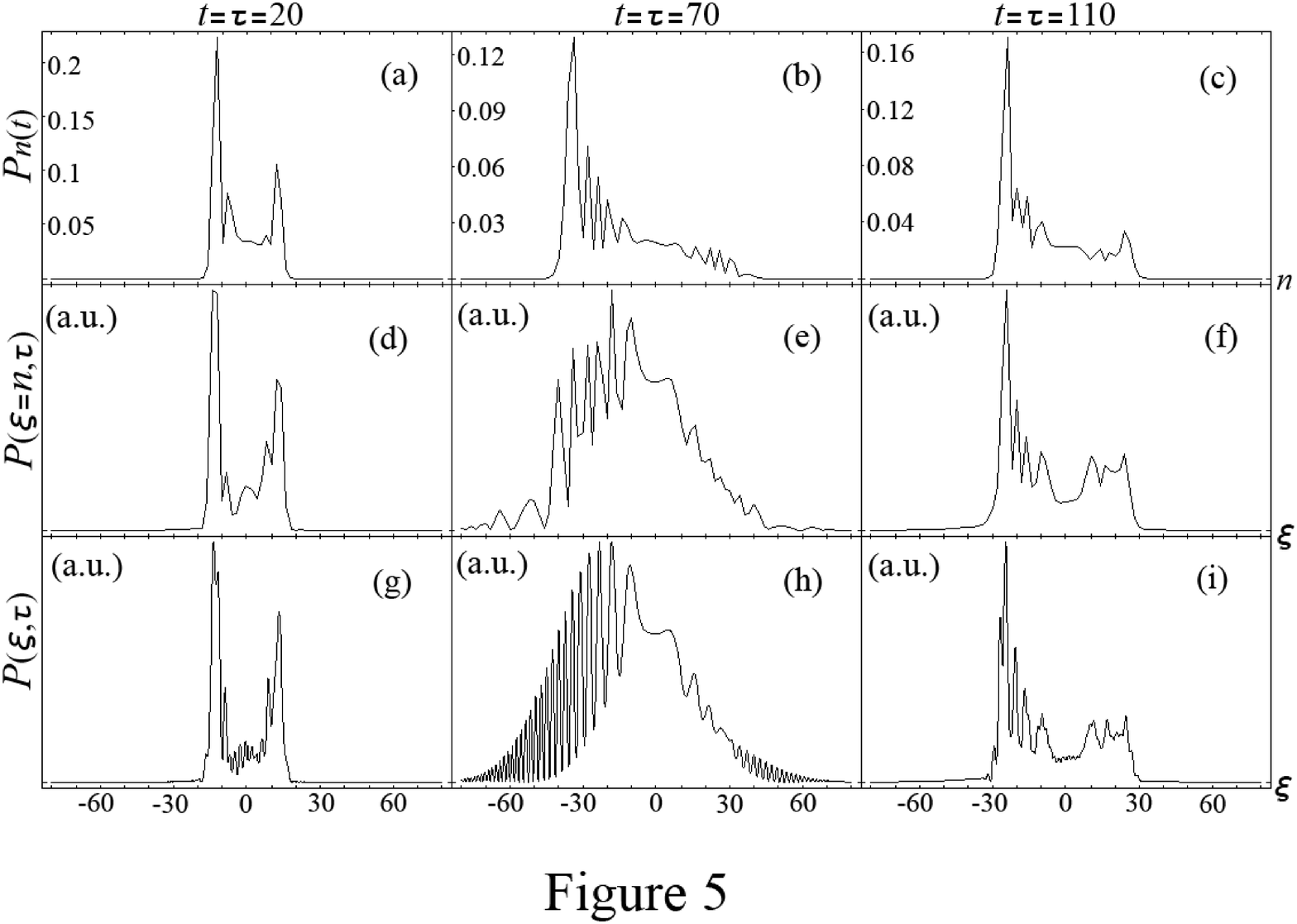}

\caption{Plots of $(n,\mathcal{P}_{n}(t))$ and $(\xi,\mathcal{P}(\xi,\tau))$
projections for three different time values: $t=20$, $t=70$, $t=110$.}
\end{figure}

Finally, we compare the evolution of the quadratic deviation $\sigma^{2}$
in position using both the exact distribution $\mathcal{P}_{n}(t)$
and the continuous distribution $\mathcal{P}(\xi,\tau)$ in Fig. 6.
We continue in the dynamic localization case with $\phi_{0}=$ $2\pi/150$,
and compare the exact case (a) with five continuous limit cases (b)
corresponding to $w=0.45$, $w=0.55$, $w=0.65$, $w=0.75$ and $w=0.85$.
Notice that the behaviors of both the exact case and the continuous
limit are similar, except for the fact that with the continuous limit
one obtains an {}``excess of quadratic deviation'', specially within
the proximity of the semiperiod, because of the already mentioned
problem with the width. %
\begin{figure}
\includegraphics[%
  scale=0.5]{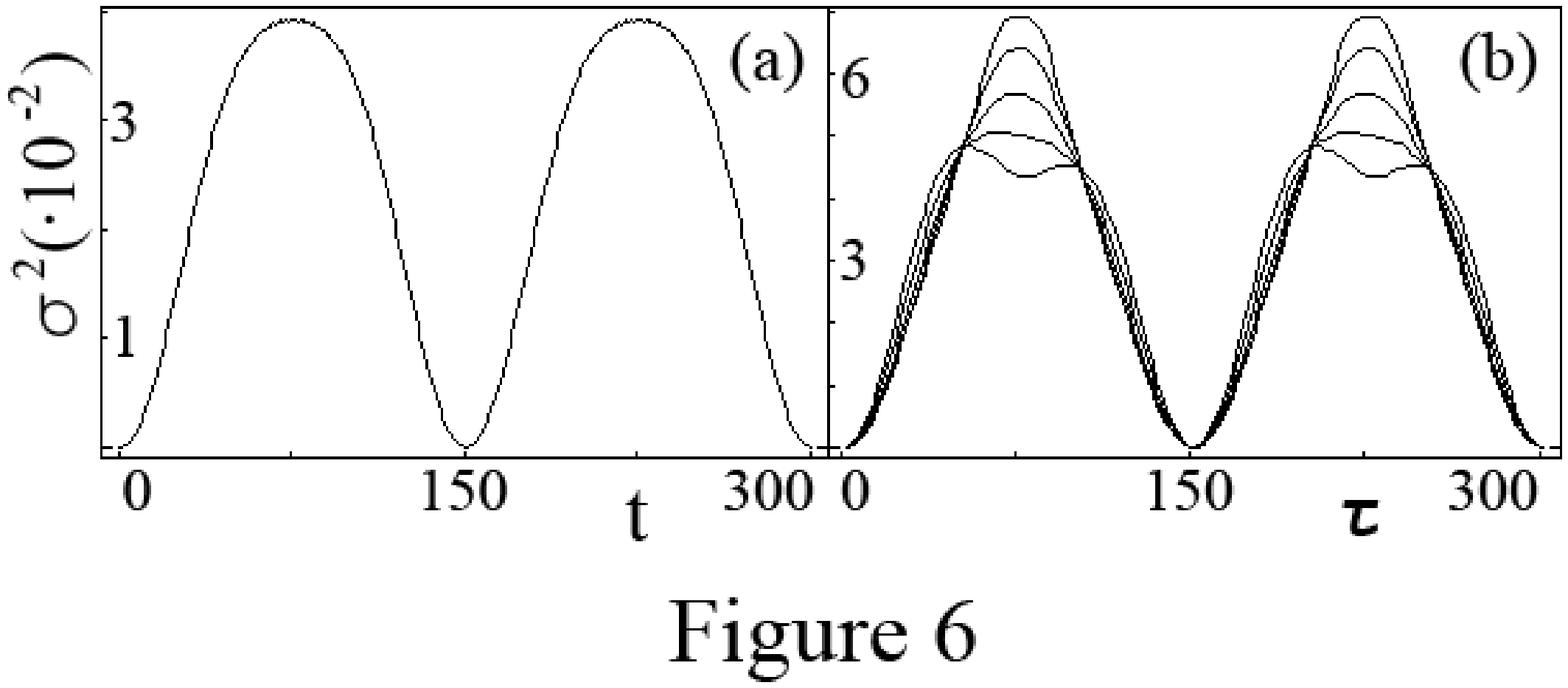}

\caption{A plot of the quadratic deviation $\sigma^{2}$ in position using
both the exact distribution $\mathcal{P}_{n}(t)$ (left) and the continuous
distribution $\mathcal{P}(\xi,\tau)$ (right), again with $\phi_{0}=$
$2\pi/150$. On the right panel, the different curves correspond to
$w=0.45$, $w=0.55$, $w=0.65$, $w=0.75$ and $w=0.85$.}
\end{figure}

The above results show that the continuous long--wavelength approximation,
Eq. (\ref{difer1}), is a good qualitative approximation, even a reasonably
good quantitative approximation, to the GQW in the dynamical localization
regime. We have already commented that the failure in describing the
diffusive dynamics occurring in the GQW for rational values of $\phi_{0}$
is due to the neglect of the third order time derivative in deriving
Eq. (\ref{difer1}), an approximation made in order to obtain analytical
expressions.

Eq. (\ref{difer1}) is a linear wave equation describing the propagation
of waves in a medium with special dispersion properties: The dispersion
coefficients (those multiplying the higher order spatial derivatives)
are time periodic, as well as the wave group velocity (the coefficient
multiplying the first order spatial derivative). Thus, the essential
for dynamical localization correspond to the vanishing of the time-averaged
group velocity, while its periodic time dependence is the responsible
for the \char`\"{}bouncing\char`\"{} of the probability distribution,
see Fig. 4. As for the rest of spatial derivatives, they introduce
a distortion on the probability distribution (due to dispersion) that
turns out to be reversible again because of the time periodicity of
the coefficients. Certainly, a group velocity that changes its sign
periodically is not a common situation for waves, but the analogy
that Eq. (\ref{difer1}) establishes provides an alternative physical
picture that, as we have seen, helps to understand dynamical localization
in the GQW and could help for the search of propagation phenomena
in which this phenomenon could manifest. In this sense, it is interesting
to notice the similarity between this equation and that describing
beam propagation in waveguides with a bent axis \cite{Longhi03},
an optical process in which Bloch oscillations and dynamical localization
have been recently experimentally observed \cite{Longhi05} (we note
that in \cite{Wojcik} the connection between the GQW and Bloch oscillations
was put forward).

\section{Conclusions}

In this article we have introduced QWs with time dependent coins.
We have considered a particularly simple case that turns out to be
equivalent to the generalized QW (GQW) introduced by Wojcik et al.
\cite{Wojcik}. This GQW exhibits very striking dynamical properties,
particularly dynamical localization. We have shown how our alternative
proposal can be used as a control mechanism. In addition, this time-dependent
QW is particularly interesting from the implementation point of view,
as only simple actions on the coin--qubit are required for that. 

We have also obtained a long--wavelength continuous approximation
of the GQW equations that have allowed us the derivation of an approximated
explicit continuous solution that works quite well during the dynamic
localization regime. The continuous equation from which this solution
was derived is a linear partial differential equation describing pulse
propagation in a dispersive medium with periodical time dependence
in the dispersion coefficients. This continuous limit has lead us
to interpret the main feature of GQW, the dynamic localization, as
a propagating solution in the dispersive medium with null mean value
of its group velocity. 

\textbf{Acknowledgments}

This work has been financially supported by Spanish Ministerio de
Educación y Ciencia and European Union FEDER through project FIS2005-07931-C03-01,
and by Grants FPA2002-00612, AYA2004-08067-C03, FPA2005-00711 and
GV05/264.

\section{Appendix}

Now we perform the long-wavelength approximation that consists in
retaining terms up to $k=1$ in Eq. (\ref{todos}), i.e.\begin{eqnarray}
\left[\frac{\partial}{\partial\tau}+\frac{1}{3!}\frac{\partial^{3}}{\partial\tau^{3}}\right]A^{\pm}\left(\xi,\tau\right) & = & \mp\sqrt{\rho}\cos\left(\phi_{0}\tau\right)\left[\frac{\partial}{\partial\xi}+\frac{1}{3!}\frac{\partial^{3}}{\partial\xi^{3}}\right]A^{\pm}\left(\xi,\tau\right)\notag\\
 &  & \mp i\sqrt{\rho}\sin\left(\phi_{0}\tau\right)\left[1+\frac{1}{2!}\frac{\partial^{2}}{\partial\xi^{2}}\right]A^{\pm}\left(\xi,\tau\right).\label{difer1-bis}\end{eqnarray}

The third-order derivative on the left--hand side makes it hard to
obtain an analytical solution. In the case of the standard QW, the
third time derivative was approximated by making use of the lowest
order expansion ($k=0$ in Eq.(\ref{todos})) \cite{Knight03}, but
in our case the time--dependent coefficient of the remaining linear
term renders this approach useless. We then make a further approximation
and neglect the third order derivative in time. By making the change

\begin{equation}
B^{\pm}\left(\xi,\tau\right)=A^{\pm}(\xi,\tau)\exp\left[\mp i\frac{\sqrt{\rho}}{\phi_{0}}\cos(\phi_{0}\tau)\right],\label{change-bis}\end{equation}
one obtains\begin{equation}
\frac{\partial}{\partial\tau}B^{\pm}\left(\xi,\tau\right)=\mp\sqrt{\rho}\left[\cos\left(\phi_{0}\tau\right)\frac{\partial}{\partial\xi}+\frac{i}{2}\sin\left(\phi_{0}\tau\right)\frac{\partial^{2}}{\partial\xi^{2}}+\frac{1}{6}\cos\left(\phi_{0}\tau\right)\frac{\partial^{3}}{\partial\xi^{3}}\right]B^{\pm}\left(\xi,\tau\right),\end{equation}
which we pass to solve here.

By Fourier transforming Eq. (\ref{difer1-bis}) one easily gets \begin{eqnarray}
B^{\pm}\left(\xi,\tau\right) & = & \int_{-\infty}^{+\infty}dk~\mathcal{B}^{\pm}\left(k,0\right)e^{ik\xi}e^{\mp i\sqrt{\rho}g\left(k,\tau\right)},\label{integral}\\
g\left(k,\tau\right) & = & \frac{\sin\left(\phi_{0}\tau\right)}{\phi_{0}}k+\frac{\cos\left(\phi_{0}\tau\right)-1}{2\phi_{0}}k^{2}-\frac{\sin\left(\phi_{0}\tau\right)}{6\phi_{0}}k^{3},\end{eqnarray}
where\begin{equation}
\mathcal{B}^{\pm}(k,0)=\frac{1}{2\pi}\int_{-\infty}^{+\infty}d\xi\
 B^{\pm}(\xi,0)e^{-ik\xi}.\end{equation}

In order to solve the integral (\ref{integral}), one must fix $B^{\pm}(\xi,0)$,
i.e., $A^{\pm}\left(\xi,0\right)$. Following \cite{Knight03}, we
assume that\begin{equation}
A_{n}^{+}\left(1\right)-A_{n}^{-}\left(1\right)\simeq A_{n}^{+}\left(0\right)+A_{n}^{-}\left(0\right),\end{equation}
 and then, by using Eq. (\ref{plusminus}),\begin{equation}
A_{n}^{\pm}(0)=\frac{1}{2}[a_{n}(0)\pm a_{n}(1)].\end{equation}
 Notice that $a_{n}(1)$ is evaluated from Eqs. (\ref{V1}, \ref{V2})
once the initial condition $a_{n}(0)$ has been fixed.

Here we consider, as usual, that the walker is initially located at
the origin of the lattice, i.e., $a_{n}(0)=0$ $\forall n$ except
for $n=0$. Then in the continuous limit we take\begin{equation}
A^{\pm}(\xi,0)=a_{0}(0)G_{0}(\xi)\pm a_{1}(1)G_{1}(\xi)\pm a_{-1}(1)G_{-1}(\xi),\label{ini}\end{equation}
 where

\begin{equation}
G_{m}(\xi)=\mathcal{N}\exp\left[-\frac{(\xi-m)^{2}}{4w^{2}}\right],\end{equation}
 with $\mathcal{N}$ a normalization constant that will be omitted
in the following. As in \cite{Knight03} we are assuming that Eq.
(\ref{todos}) is correct only for the long--wavelength components
by taking an initial condition that {}``smears out'' the lower--wavelength
components.

Now, by using Eqs. (\ref{change-bis}) and (\ref{ini}), one easily
obtains\begin{equation}
\mathcal{B}^{\pm}(k,0)=\left[a_{0}(0)\pm\sum_{m=\pm1}a_{m}(1)\exp\left(mik\right)\right]e^{-\left(w^{2}k^{2}\pm i\sqrt{\rho}/\phi_{0}\right)},\end{equation}
and with this, the result of (\ref{integral}) reads\begin{equation}
B^{\pm}\left(\xi,\tau\right)=\left[a_{0}(0)\mathcal{Z}^{\pm}\left[\pm\xi,\tau\right]\pm\sum_{m=\pm1}a_{m}(1)\mathcal{Z}^{\pm}\left[\pm\left(\xi-m\right),\tau\right]\right]e^{\mp i\sqrt{\rho}/\phi_{0}},\label{B}\end{equation}
where the functions $Z^{\pm}(\xi^{\prime},\tau)$ are\begin{eqnarray}
\mathcal{Z}^{\pm}(\xi^{\prime},\tau) & = & \int_{-\infty}^{+\infty}dq\exp\left[i\alpha q-\frac{i}{3}\beta q^{3}-(1\pm i\gamma)q^{2}\right],\label{Z}\\
\text{\thinspace\thinspace}\alpha & = & \frac{\xi^{\prime}}{w}-\frac{\sqrt{\rho}}{\phi_{0}w}\sin(\phi_{0}\tau),\notag\\
\beta & = & -\frac{\sqrt{\rho}}{2\phi_{0}w^{3}}\sin(\phi_{0}\tau),\notag\\
\gamma & = & \frac{\sqrt{\rho}}{2\phi_{0}w^{2}}[\cos(\phi_{0}\tau)-1].\notag\end{eqnarray}
Their solutions read \cite{Miyagi79,Knight03}\begin{eqnarray}
\mathcal{Z}^{\pm}(\xi^{\prime},\tau) & = & \frac{1}{|\beta|^{1/3}}\mathcal{A}_{i}\left(a\right)e^{b},\label{Zsol}\\
a & = & \frac{1-\alpha\beta-\gamma^{2}\pm2i\gamma}{|\beta|^{4/3}},\notag\\
b & = & \frac{2-3\alpha\beta-6\gamma^{2}}{3\beta^{2}}\mp i\gamma\frac{3\alpha\beta+2\gamma^{2}-6}{3\beta^{2}},\notag\end{eqnarray}
where $\mathcal{A}_{i}(z)$ is the Airy function.

Finally, by using Eqs. (\ref{B}) and (\ref{change-bis}), we can
write down the solution for the fields $U^{\pm}(\xi,\tau)$ and $D^{\pm}(\xi,\tau)$\begin{eqnarray}
U^{\pm}\left(\xi,\tau\right) & = & \left\{ u_{0}\left(0\right)\mathcal{Z}^{\pm}\left[\pm\xi,\tau\right]\pm\sum_{m=\pm1}u_{m}(1)\mathcal{Z}^{\pm}\left[\pm\left(\xi-m\right),\tau\right]\right\} e^{\pm2iw^{2}\gamma}\\
D^{\pm}\left(\xi,\tau\right) & = & \left\{ d_{0}\left(0\right)\mathcal{Z}^{\pm}\left[\pm\xi,\tau\right]\pm\sum_{m=\pm1}d_{m}(1)\mathcal{Z}^{\pm}\left[\pm\left(\xi-m\right),\tau\right]\right\} e^{\pm2iw^{2}\gamma}\end{eqnarray}
The total probability of finding the walker in the position $\xi$
at time $\tau$ can be easily calculated as\begin{eqnarray}
\mathcal{P}(\xi,\tau) & = & \mathcal{P}^{u}(\xi,\tau)+\mathcal{P}^{d}(\xi,\tau)\label{conprob}\\
\mathcal{P}^{u}(\xi,\tau) & = & |u(\xi,\tau)|^{2}=\left|U_{n}^{+}\left(t\right)+\left(-1\right)^{t}U_{n}^{-}\left(t\right)\right|^{2},\,\,\,\,\,\,\notag\\
\mathcal{P}^{d}(\xi,\tau) & = & |d(\xi,\tau)|^{2}=\left|D_{n}^{+}\left(t\right)+\left(-1\right)^{t}D_{n}^{-}\left(t\right)\right|^{2}.\notag\end{eqnarray}

\end{document}